\documentclass[twocolumn, superscriptaddress, pra]{revtex4}

\usepackage{graphicx, epsfig, fancybox}
\usepackage{amssymb,amsmath}

\def \hf{\tfrac{1}{2}}    

\def \ord{\mathcal{O}}

\def\lbc{\left[}    \def\rbc{\right]}

\newcommand{\ket}[1]{\left|{#1}\right.\rangle}
\DeclareMathOperator{\tr}{tr}

\begin{document}

\title{Entanglement and level crossings in frustrated ferromagnetic rings}

\author{Masudul Haque}

\affiliation{Max-Planck Institute for the Physics of Complex
Systems, N\"othnitzer Str.~38, 01187 Dresden, Germany}

\author{V.~Ravi Chandra}

\affiliation{Max-Planck Institute for the Physics of Complex
Systems, N\"othnitzer Str.~38, 01187 Dresden, Germany}

\author{Jayendra N.~Bandyopadhyay}

\affiliation{Max-Planck Institute for the Physics of Complex
Systems, N\"othnitzer Str.~38, 01187 Dresden, Germany}

\affiliation{Department of Physics, National University of Singapore,
Singapore 117542}

\begin{abstract}

We study the entanglement content of a class of mesoscopic tunable magnetic
systems.  The systems are closed finite spin-1/2 chains with ferromagnetic
nearest-neighbor interactions frustrated by antiferromagnetic
next-nearest-neighbor interactions.  The finite chains display a series of
level crossings reflecting the incommensurate physics of the corresponding
infinite-size chain.  We present dramatic entanglement signatures
characterizing these unusual level crossings.  We focus on multi-spin and
global measures of entanglement rather than only one-spin or two-spin
entanglements.  We compare and contrast the information obtained from these
measures to that obtained from traditional condensed matter measures such as
correlation functions.

\end{abstract}

\maketitle

\section{Introduction}

The study of entanglement measures in the eigenstates of condensed matter
Hamiltonians is a rapidly growing field of interdisciplinary research
\cite{Amico-et-al_RMP08}.  While a great many results have been reported, the
typical situation is that entanglement studies of condensed matter systems
provide alternative signatures of phenomena already known from more
traditional condensed-matter techniques.
In this article, we use measures from quantum information to
characterize a spin system which is poorly understood in the condensed
matter literature, namely, spin rings with ferromagnetic
nearest-neighbor interactions frustrated by antiferromagnetic
next-nearest-neighbor interactions.
We show how the unusual properties of this system are reflected in quantum
information quantities.

The systems we present are closed rings of $N\sim\ord(10)$ spins.  The study
of finite-size quantum systems have enjoyed a huge resurgence in the past
decade because of cold-atom and nanoscience developments, which provide the
context for designing and studying finite-size Hamiltonians for their own
fundamental properties rather than as mere tools for investigating the
thermodynamic limit $N\rightarrow\infty$.  We will focus on a sequence of
ground-state level crossings in such systems, which reflect
lattice-incommensurate correlations.  
We show that certain measures of entanglement
%as well as the ``fidelity'' of the ground-state wavefunction, 
display remarkably strong parameter dependences near these level crossings.
Our calculations of the \emph{concurrence} show that the entanglement
structure of the states between successive level crossings is ideal for tuning
the distance between spins which are mutually entangled.
We also present some results on entanglement \emph{dynamics}, again focusing
on the effect of the level crossings.

% The spin rings we study display a series of level crossings.  
It is natural to think of ground state level crossings as finite-size analogs
of first-order phase transitions in the thermodynamic limit.  As a corollary,
one expects that ground state properties change suddenly at the crossing
(`transition') point and that away from the exact crossing point the system
possesses no knowledge of the proximity of the crossing.  In this regard, our
quantum information quantities reveal a peculiarity of the crossings in our
spin rings, because they seem to evolve in response to the proximity of the
level crossings, already at some distance away from the parameter value where
the crossing occurs, \emph{i.e.}, the ground state is ``aware'' of the nearby
crossing.  This counter-intuitive phenomenon is clearly seen in multi-spin
(but bipartite) entanglements as well as fidelities, but not in traditional
condensed-matter quantities such as correlation functions.
Note that, since the level crossings are not expected in the thermodynamic
limit, this issue is meaningful only in our mesoscopic context.

To calculate a bipartite entanglement, one has to first decide on a partition.
For a condensed-matter model or even its finite-size versions, there are a
large number of ways in which the system can be bi-partitioned.  A variety of
different partitionings, and correspondingly different entanglements, have
been used for entanglement studies in condensed matter models.  
Many studies have focused on the entanglement between two sites as measured by
the concurrence or negativity (\emph{e.g.}, \cite{OsterlohFazio_Nature02,
OsborneNielsen_PRA02, WangZanardi_PhysLett02, BoseChattopadhyay_PRA02,
Syljuasen_PRA03, Subrahmanyam_PRA04, ChoMcKenzie_PRA06}), or the entanglement
between one spin/site with the rest (\emph{e.g.}, \cite{TribediBose_PRA07,
AlipourKarimipourMemarzadeh_PRA07, KoppLeHur_PRL07}).  Such measures
generically are not expected to provide information different from traditional
condensed matter quantities like 2-point correlation functions, since the
entanglement measure can be related easily to correlation functions
(\emph{e.g.}, \cite{WangZanardi_PhysLett02, Syljuasen_PRA03,
Subrahmanyam_PRA04, ChoMcKenzie_PRA06, KoppLeHur_PRL07}).
In this sense, many-site measures are more innovative.  Examples are block
entanglement in 1D \cite{block_1D} and in 2D \cite{block_2D}, and more
recently studied spatially disconnected partitions such as sublattice
entanglement \cite{sublattice-entanglement} and entanglement between labeled
itinerant particles \cite{particle-entanglement}.  Such entanglement measures
are not easily related to usual correlation functions, and therefore hold the
possibility of providing genuinely new tools for probing collective phenomena
in many-particle systems.
For our finite-sized chains, we will present entanglement results for various
partitions, both two-site and many-site.

The dynamics of entanglement measures in many-particle systems has attracted
some attention recently \cite{entanglement-dynamics}, but is still a
relatively poorly understood topic, reflecting the general difficulty of
non-equilibrium physics in quantum condensed matter.
Entanglement dynamics has also gained interest from the quantum information
perspective, \emph{e.g.}, in the context of adiabatic quantum computing
\cite{AdiabaticQC} and in the context of quantum state transfer (`quantum
communication') through spin chains \cite{SpinChain_QuantumCommunication}.
Obviously, the number of different conceivable non-equilibrium situations is
large; we are forced to restrict the types of dynamics considered.  In this
article we will describe the entanglement dynamics after a sharp change
(\emph{quench}) of interaction parameter.

After describing in Sec.~\ref{sec_partitns-n-entanglements} the various
entanglement measures and partitions to be used, in Sec.~\ref{sec_model} we
introduce the Hamiltonian and its condensed-matter context.
Sec.~\ref{sec_model} also describes the level-crossings and compares to other
examples of similar crossings known to us.  The next sections present
entanglement results, starting in Sec.~\ref{sec_corrfn-concurr} with the
quantities containing the most traditional information, namely, the
concurrences which are readily related to correlation functions.  Sections
\ref{sec_multispin_derivatives} and \ref{sec_fidelities} present
quantum-information quantities which are more distinct from condensed-matter
measures.  Sec.~\ref{sec_dynamics} describes our results on entanglement
dynamics.

%% Section \ref{sec_multispin_derivatives} presents more `multi-spin'
%% entanglements and their derivatives, and Section \ref{sec_fidelities}
%% describes ground-state fidelities.

\section{Partitions and entanglements}  \label{sec_partitns-n-entanglements}

\begin{figure}
\centering
\includegraphics[width=0.8\columnwidth]{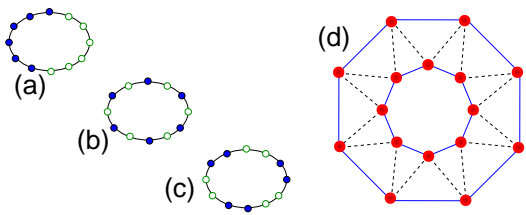}
\caption{\label{fig_partition-cartoons} 
(a-c) Various partitions of closed chain for calculating bipartite
  entanglement entropy.  Spins belonging to partition $A$ ($B$) are
  represented by open (full) circles.  (d) Geometry for exploiting the
  sublattice entanglement $S_{\rm s.l.}$, \emph{e.g.}, in an experimental
  realization.  In this setup $S_{\rm s.l.}$ is simply the entanglement between
  inner and outer rings.  Dashed and full lines represent $J_1$ and $J_2$
  interactions respectively.
}
\end{figure}

We will mostly measure entanglement using the \emph{entanglement entropy}.
The entanglement entropy between partitions $A$ and $B$ is defined using the
reduced density matrix of one part (e.g., $\rho_A = \tr_B\rho$ obtained by
tracing over $B$ degrees of freedom) to calculate the von Neumann entropy,
$S_A = -\tr[\rho_A\ln\rho_A]$.
Since the purpose of using entanglement in many-particle systems is to provide
viewpoints not present in usual condensed-matter measures, we want to consider 
more `non-local' entanglements.  The partitions we choose for this purpose are
shown in Figure \ref{fig_partition-cartoons}.  

The most obvious choice is for $A$ and $B$ to be spatially connected blocks,
in particular for each partition to be a connected half of the ring (Fig.\
\ref{fig_partition-cartoons}a).  This partitioning is suggested by many
existing results on block entanglement entropies in one dimension
\cite{block_1D}.

Another choice is to take all the odd sites as partition $A$ and the even
sites as partition $B$ --- this is then the entanglement between $A$ and $B$
\emph{sublattices}, in usual quantum magnetism language (Fig.\
\ref{fig_partition-cartoons}b).  The entanglement between sublattices has been
studied for a number of spin chains and lattices, and extrema of $S_A$ have
been claimed to characterize some quantum phase transitions
\cite{sublattice-entanglement}.  Note also that since it is natural to think
of the $J_1$-$J_2$ chain as a `zigzag ladder' (\emph{e.g.},
\cite{Hamada-et-al_JPSJ88, Xiang_PRB06,
ChhajlanyRichter_MG-entanglement_PRA07}), the sublattice entanglement is
simply the entanglement between the two legs of such a ladder.  In the
mesoscopic (finite-size) context that we are interested in, a $J_1$-$J_2$ ring
is likely to be implemented as shown in Fig.\ \ref{fig_partition-cartoons}d;
the two `legs' are now two $J_2$ rings (either inner or outer rings, or an
upper and a lower layer), connected by zigzag $J_1$ interactions.  From an
experimental quantum information processing perspective, the sublattice
entanglement is natural as it is simply the entanglement between inner and
outer (or upper and lower) rings.

Since the antiferromagnetic interaction in our case is the
\emph{next}-nearest-neighbor interaction, we extend the idea of the sublattice
partition ($ABAB...$) and also use the $AABBAA...$ partition shown in Fig.\
\ref{fig_partition-cartoons}c.

We call the entanglement entropies corresponding to the above three partitions
respectively $S_{1/2}$, $S_{\rm s.l.}$, $S_{2/4}$.

Another nonlocal quantity motivated by quantum information theory is the
fidelity, the overlap between ground state wavefunctions at slightly different
system parameters \cite{fidelity-in-condmat}.
%
%% This quantity has been proposed as a useful indicator of quantum phase
%% transitions \cite{fidelity-in-condmat}.
%
We describe fidelities in our spin rings in Section \ref{sec_fidelities}.

We also present results for more `local' measures of entanglement, namely, 
 (1) entanglement entropy for 2-site partitions, and (2) 
the concurrence $C_r$ measuring the entanglement between sites $i$ and $i+r$
in the environment provided by the rest of the chain
\cite{Wooters_concurrence_PRL98}.  The concurrence is defined as
\[
C_r = \max\Big[\;0,\; \Big(\sqrt{\lambda_1} -\sqrt{\lambda_2}
  -\sqrt{\lambda_3} -\sqrt{\lambda_4}\Big) \Big] \; ,
\]
where $\lambda_i$ are the eigenvalues in decreasing order of the matrix
$\rho_A (\sigma_y\otimes\sigma_y) \rho_A^* (\sigma_y\otimes\sigma_y)$, and
$\rho_A$ is the reduced density matrix of the two-site subsystem. 
Since we will consider singlet states, two-site reduced density matrices are
strongly constrained, and the concurrence is a simple function of the
spin-spin correlation functions.
Two spins are entangled ($C_{r}>0$) if the correlations between them are
sufficiently antiferromagnetic, $\langle{S^z_i}S^z_{i+r}\rangle<-1/12$, and in
that case $C_{r} = -6\langle{S^z_0}S^z_r\rangle-\hf$ (derived in, \emph{e.g.},
\cite{Subrahmanyam_PRA04}; see also Ref.~22 in
\cite{ChhajlanyRichter_MG-entanglement_PRA07}).
The behavior of concurrence in various spin models is briefly reviewed from
the literature in Sec.~\ref{sec_corrfn-concurr} in order to provide context
for our concurrence results.

\section{Frustrated ferromagnetic rings}  \label{sec_model}

We are concerned with closed chains of $N$ localized spin-1/2 objects,
\emph{i.e.}, a finite-size spin chain, interacting via the Hamiltonian
\begin{equation}  \label{eq_Hamilt}
H = J_1 \sum_{i}  {\bf S}_i \cdot {\bf S}_{i+1} +  
J_2 \sum_{i} {\bf S}_i \cdot {\bf S}_{i+2}  \; ,
\end{equation}
where $J_1<0$ $J_2>0$, and $i$ is the site index obeying periodic boundary
conditions appropriate to a ring.
%  ($i=N+1$ is identified with $i=1$.)  
We define $\beta = J_2/|J_1|$ and $\bar{\beta}=\beta^{-1}$.  This spin chain
is known for any even $N$ to have a singlet ground state for $\bar{\beta}<4$
and a $(N+1)$-fold degenerate ferromagnetic ground state manifold for
$\bar{\beta}>4$ \cite{TonegawaHarada_JPSJ89}.  At $\bar{\beta}=4$ a singlet
ground state is degenerate with the ferromagnetic manifold; these degenerate
states are known exactly \cite{Hamada-et-al_JPSJ88, TonegawaHarada_JPSJ89,
Dmitriev-etal_PRB97, Dmitriev-etal_PRB00, SuzukiTakano_cm08}.
%
%% Relatively little is known about the singlet ground states for
%% $\bar{\beta}\in(0,4)$.

We will confine ourselves to the singlet region, $\bar{\beta}\in(0,4)$.  The
numerical results are presented for ring sizes divisible by 4, in order to
avoid complications due to the parity of $N/2$.  The rings for other even
sizes have very similar behavior.

Although far less studied compared to the $J_1>0$ case (Majumdar-Ghosh chain),
the frustrated ferromagnet Hamiltonian has already been known for some time to
possess some peculiar properties \cite{Hamada-et-al_JPSJ88,
TonegawaHarada_JPSJ89, Chubukov_PRB1991, NersesyanGogolinEssler_PRL98,
CabraHoneckerPujol_EPJB00, ItoiQin_PRB01}.
More recently, this model has attracted a fresh surge of theoretical attention
\cite{HoneckerVekua_PRB06, VekuaHonecker_PRB07, Xiang_PRB06,
DmitrievKrivnov_PRB06, KeckeMomoiFurusaki_PRB07, RichterIhleDrechsler_PRB08,
HikiharaKeckeMomoiFurusaki_cm08, Laeuchli_cm08, Mahdavifar_cm08}, partly due
to the appearance of materials whose spin physics is described by this model
(see references in \cite{HoneckerVekua_PRB06, Xiang_PRB06}).  Most studies
concentrate on finite magnetic fields or finite temperatures and on properties
in the thermodynamic limit \cite{Chubukov_PRB1991,
NersesyanGogolinEssler_PRL98, CabraHoneckerPujol_EPJB00, ItoiQin_PRB01,
HoneckerVekua_PRB06, VekuaHonecker_PRB07, Xiang_PRB06, DmitrievKrivnov_PRB06,
KeckeMomoiFurusaki_PRB07, RichterIhleDrechsler_PRB08,
HikiharaKeckeMomoiFurusaki_cm08, Laeuchli_cm08}, or very near the transition
point at $\bar{\beta}=4$ \cite{KrivnovOvchinikov_PRB96, DmitrievKrivnov_PRB06,
SuzukiTakano_cm08, Mahdavifar_cm08}.

%(more citations? maybe from Hikihara-Furusaki paper cm08? or Sudan's list?)

\emph{Proliferation of ground state level crossings} ---
An unusual feature of the singlet region $\bar{\beta} \in(0,4)$ is the number
of ground state level crossings, which proliferate as we increase the number
of spins \cite{TonegawaHarada_JPSJ89}.  For $N=$ 8, 12, 16, 20,... spins,
there are respectively one, two, three, four,... level crossings in this
region.
A very likely interpretation is that the natural correlations in the
chain for  $\bar{\beta}\in(0,4)$ are incommensurate with
wavevector dependent on $\beta$.  This incommensurability is resolved
in different ways by the different ground states.  For larger rings,
there are a greater number of possibilities to accommodate the
incommensurate correlations into the finite system, and hence the
larger number of level crossings.
The ground state lattice momentum alternates between $0$ and $\pi$ in the
regions between successive level crossings. 

Very similar level crossings appear in other spin models known or thought to
have spiral correlations.  The examples known to us are
(1) the Majumdar-Ghosh chain for $J_2>0.5J_1$
\cite{TonegawaHarada_JPSJ87}; 
(2) the `sawtooth' chain of Ref.\ \cite{Ravi_sawtooth_PRB04};
(3) the frustrated ferrimagnetic chain of Ref.\
    \cite{IvanovRichterSchollwoeck}. 
This phenomenon of proliferating level crossings
in finite-size rings as a result of incommensurate correlations, however, is
not widely known or discussed in the condensed matter literature.

%% While our work provides some clues about the thermodynamic limit also, in this
%% paper we will study entanglement signatures of these mesoscopic-system level
%% crossings themselves.

\section{Correlation functions and concurrences} \label{sec_corrfn-concurr}

\begin{figure}
\centering
 \includegraphics[width=0.95\columnwidth]{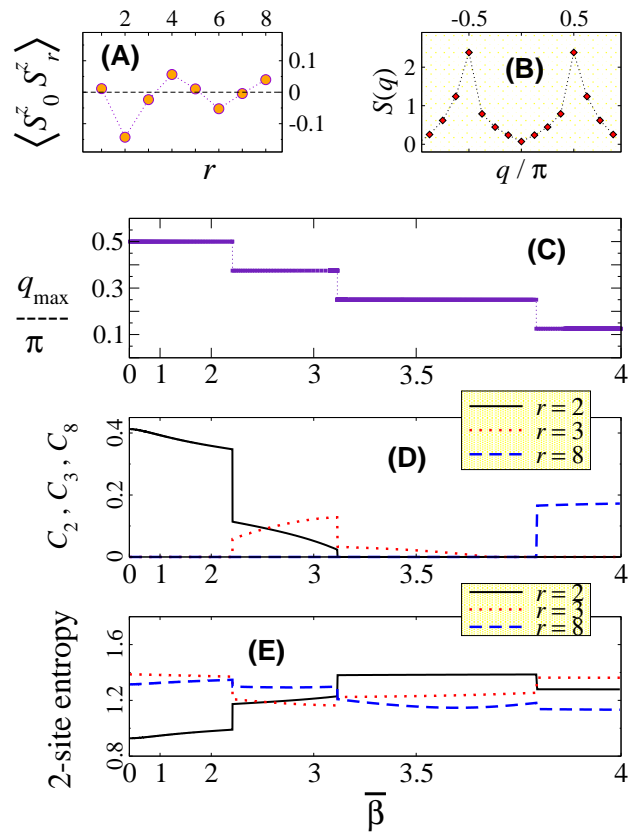}
\caption{\label{fig_StrctureFactor-n-qmax} 
Correlation functions and related quantities for $N=16$.  
(A) correlation function at $\bar{\beta}=2$.    
(B) The structure factor, $S(q)$,  at $\bar{\beta}=2$.
(C) Position of the maximum of $S(q)$,  as a function of $\bar{\beta}$.
(D) Concurrences between two spins at distance $r =$ 2, 3, 8.
(E) Entropy of entanglement between partition consisting of sites ($i$,
$i+r$), and the remaining $N-2$ sites.  Shown also for $r =$ 2, 3, 8.
In (C-E), nonlinear horizontal scale highlights the region near $\bar{\beta}=
4$. 
}
\end{figure}

In this section we present more local quantities, in particular traditional
condensed matter quantities like spin-spin correlation functions and structure
factors, and closely related entanglements like concurrences.

For singlet states, the correlation function
$\langle{\bf{S}_i}\cdot\bf{S}_j\rangle$ is three times
$\langle{S^z_i}S^z_j\rangle$; we will discuss the latter as a function of
$r=|i-j|$.  Between $\bar{\beta}=0$ and the first crossing, the signs of this
quantity has the structure
\[
+ \, , -  \, , -  \, , +  \, , +  \, , -  \, , -  \, ,  +  \, , +  \, , ....
\]
for $r=1, 2, 3, 4, ....$, indicating that the state in this region has a
`spin-density-wave'-like
$\ket{\uparrow\uparrow\downarrow\downarrow\uparrow\uparrow\ldots}$ structure.
An example is shown in Fig.\ \ref{fig_StrctureFactor-n-qmax}A for
$N=16$.
The oscillating magnitude/sign pattern also suggests `nematic'-like
correlations \cite{MomoiShannon_PTPS05, ShannonMomoiSindzingre_PRL06,
VekuaHonecker_PRB07}; detailed distinctions are inappropriate since this is
not a macroscopic phase.

The sign pattern of $\langle{S^z_0}S^z_r\rangle$ changes at each level
crossing; for $N=16$ the patterns in the four regions (separated by level
crossings) are 
\begin{gather*}
+ \; -  \; -  \; +  \; +  \; -  \; -  \;   +  \\
+ \; -  \; -  \; -  \; +  \; +  \; -  \;   -  \\
+ \; -  \; -  \; -  \; -  \; -  \; +  \;   +  \\
+ \; +  \; +  \; -  \; -  \; -  \; -  \;   -  
\end{gather*}

The structure factor $S(q)$, which is the Fourier transform of the real-space
correlation function, has peaks at $q_{\rm max} = (2\pi/N)m$, with the integer
$m$ decreasing in unit steps at each level crossing, from $m=N/2$ at
$\bar{\beta}=0$ to $m=1$ near the ferromagnetic transition point
\cite{TonegawaHarada_JPSJ89}.  Similar behavior (Fig.\
\ref{fig_StrctureFactor-n-qmax}C) is also seen in the other
examples known to us where finite-size spin systems have level crossings due
to spiral correlations \cite{TonegawaHarada_JPSJ87, Ravi_sawtooth_PRB04}.

We now consider the concurrences $C_r$ between spins $i$ and $i+r$ on the rings.
For most simple spin models, the concurrence does not extend significantly
beyond the next-nearest-neighbor site, \emph{i.e.}, $C_{r>2}=0$ for many spin
models, such as the $XY$ model in transverse field
\cite{OsterlohFazio_Nature02}, the Heisenberg model ground states on 1D chains
and various 2D lattices \cite{Subrahmanyam_PRA04}, and even in the
Majumdar-Ghosh ($J_1>0$, $J_2>0$) chain \cite{QianShiLiSongCPSun_PRA05}.  In
the XYZ model in a magnetic field, the concurrence extends to several lattice
sites, but still has finite range \cite{Roscilde-et-al_PRL04}.  Long-range
concurrences are rare; the known examples correspond to specific phase
transitions \cite{AmicoBaroniFubini-etal_PRA06} or to carefully constructed
models and geometries \cite{CamposVenuti-etal_PRL06, CamposVenuti-etal_PRA07}.

In this context, it is noteworthy that our highly frustrated spin rings
display nonzero concurrences at all length scales, depending on the
interaction parameter $\bar{\beta}$ (Fig.\ \ref{fig_StrctureFactor-n-qmax}D).
For any particular value of $\bar{\beta}$, only a couple of concurrences are
nonzero, but not necessarily the ones with small $r$.  The $r$ values at which
$C_r$ are nonzero, increases as $\bar{\beta}$ changes from 0 to 4, from $r=2$
to $r=N/2$.
In the region near the ferromagnetic transition $\bar{\beta}=4$, there is
nonzero concurrence between spins at opposite ends of the ring
($C_{r=N/2}>0$), reflecting the unusual long-range quantum correlations of the
so-called UDRVB state \cite{Hamada-et-al_JPSJ88, TonegawaHarada_JPSJ89} near
$\bar{\beta}=4$.
The behavior we have found indicates that the interaction here acts as a knob
to tune the distance between spins with nonzero entanglement.  This tunable
distance is quite different from the `entanglement range' of Ref.\
\cite{AmicoBaroniFubini-etal_PRA06}; our concurrences are not decaying with
distance but are nonzero only at a certain distance. 

Fig.\ \ref{fig_StrctureFactor-n-qmax}E shows the entanglement between a
two-spin partition (consisting of site $i$ and site $i+r$) and the rest of the
ring.  Loosely speaking, this entanglement entropy is large (small) when the
concurrence \emph{between} the two spins is small (large).   This reflects the
intuitive idea that objects highly entangled with each other are generally
weakly entangled with their environment.

\section{Entanglements and  their derivatives} \label{sec_multispin_derivatives}

\begin{figure}
\centering
\includegraphics[width=0.95\columnwidth]{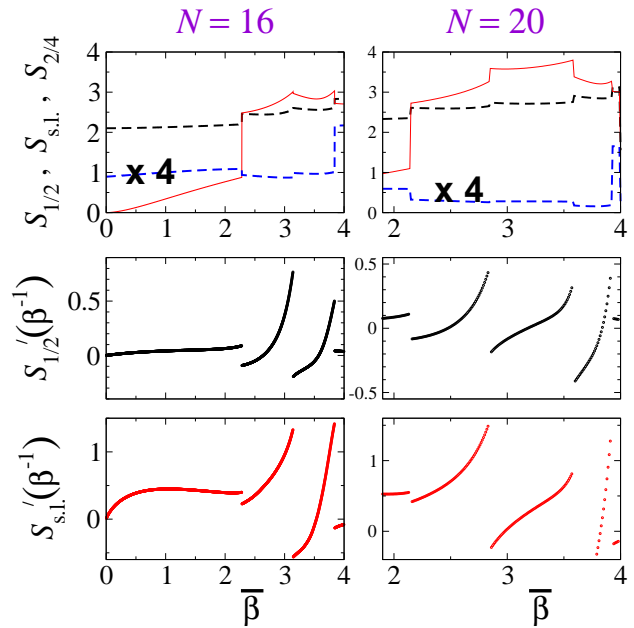}
\caption{\label{fig_entnglmnts-A} 
16-spin and 20-spin rings.  Top: half-block and sublattice entanglements,
$S_{1/2}$ (upper dashed), $S_{\rm s.l.}$ (solid), $S_{2/4}$ (lower dashed,
magnified), against $\bar{\beta}$ in the singlet region.  For the $N=20$
curves, smaller values of $\bar{\beta}$ are omitted, to focus on the level
crossings.  
Middle and bottom: derivatives of $S_{1/2}(\bar{\beta})$, $S_{\rm
s.l.}(\bar{\beta})$, respectively.
}
\end{figure}

Figure \ref{fig_entnglmnts-A} displays entanglements $S_{1/2}$,
$S_{\rm s.l.}$, $S_{2/4}$ for the partitions explained in Figure
\ref{fig_partition-cartoons}, and also the derivatives with respect to
$\bar{\beta}$ for the first two cases.

The block entanglement entropy $S_{1/2}$ is large compared to, say, the
singlet ground state of the nearest-neighbor Heisenberg ground state, in which
case $S_{1/2}\sim$1.28(1.354) for $N=$ 16(20).  This is not surprising, as the
longer-range interaction causes larger entanglement.  The Majumdar-Ghosh chain
(positive $J_1$ and $J_2$) similarly has relatively large block entanglement
\cite{ChhajlanyRichter_MG-entanglement_PRA07}.

The sublattice entanglement $S_{\rm s.l.}$ starts from zero at $\bar{\beta}=0$,
because the two sublattices are decoupled in the $J_2\rightarrow\infty$ limit.
After the first level crossing, $S_{\rm s.l.}$ also reaches larger values.
Note that $S_{\rm s.l.}$ is quite large in the Heisenberg and Majumdar-Ghosh
chains; in comparison the ground-state singlets of our system have somewhat
lower sublattice entanglement.

The derivatives of $S_{1/2}(\bar{\beta})$ and $S_{\rm s.l.}(\bar{\beta})$
(Fig.\ \ref{fig_entnglmnts-A} lower panels) are similar and have some striking
features.  First, at each level crossing except the leftmost, the derivatives
have shapes reminiscent of a resonance, although there is no real divergence.
In other words, the entanglement curves $S(\bar{\beta})$ become steep but not
completely vertical near each crossing.  Second, the $S'(\bar{\beta})$ curves
appear to respond (curve upward or downward) already some distance away from
the level crossing.  While phase transition language is not completely
appropriate for our mesoscopic arrangements, the crossings in this sense are
loosely speaking ``second-order'' like rather than ``first-order'' like.
Finally, the $S'(\bar{\beta})$ curves at the leftmost crossing lacks the
resonance-like feature, indicating that the first level crossing is different
in nature.

The $S_{2/4}$ entanglement entropy is peculiarly small (shown 4-fold magnified
in Fig.\ \ref{fig_entnglmnts-A}), although the relevant reduced density matrix
has the same dimensions as the other two partitions.  This peculiarity remains
unexplained at present.

% Other block entanglements. 

\section{Fidelities}  \label{sec_fidelities}

\begin{figure}
\centering \includegraphics[width=0.95\columnwidth]{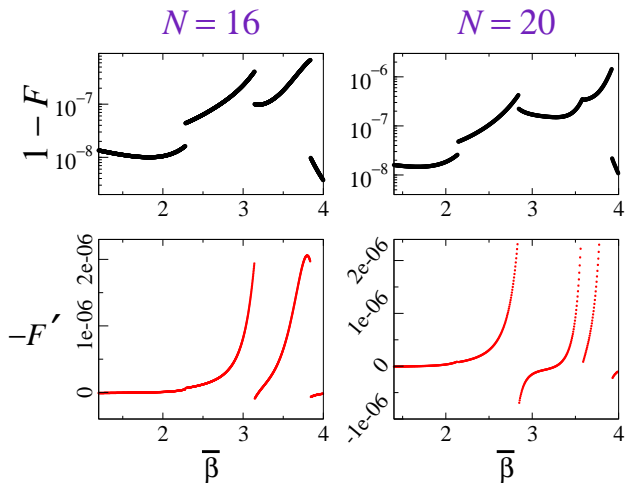}
\caption{\label{fig_fidelities} 
Fidelity $F_{\epsilon}(\bar{\beta})$ with $\epsilon=10^{-3}$, and its derivative.
Since $F$ is generally very close to unity, it is more convenient to plot
$1-F$ on a logarithmic scale.
Note that the discontinuity at the leftmost level crossing is pronounced in
the fidelity but almost invisible in its derivative.
}
\end{figure}

The ground-state fidelity is defined as the overlap between ground state
wavefunctions at nearby parameter values:
\[ 
F_{\epsilon}(\bar{\beta}) = \Big< \, \psi(\bar{\beta}-\hf\epsilon) \, \Big|
\, \psi(\bar{\beta}+\hf\epsilon) \Big>  \; .
\] 
Recent studies indicate that the fidelity provides useful signatures of
quantum phase transitions \cite{fidelity-in-condmat}.  Since we concentrate on
mesoscopic spin structures, our purpose is not to study phase transitions 
% (see however Sec.\ \ref{sec_thermodynlimit}) 
but to present fidelity signatures of the level crossings.

Figure \ref{fig_fidelities} presents fidelity data for $\epsilon=10^{-3}$.
The derivatives $F'(\bar{\beta})$ again show striking resonance-like features.
The special nature of the leftmost crossing is manifested even more
dramatically here -- the derivatives match almost perfectly across this
crossing, even though the fidelities themselves are discontinuous.

\section{Time evolution after a quench} \label{sec_dynamics}

We now turn to the \emph{dynamics} of entanglement in the frustrated
ferromagnetic chain.  While each entanglement measure has its own dynamics
when the quantum state of the spin ring evolves outside equilibrium, a full
study of the dynamics of all such measures is outside the scope of the present
investigation, especially since such comprehensive entanglement dynamics
studies are yet to be completed for simpler spin systems.  We therefore
restrict the presentation to the dynamics of the sublattice entanglement
entropy ($S_{\rm s.l.}$).

\begin{figure}
\centering
 \includegraphics[width=0.95\columnwidth]{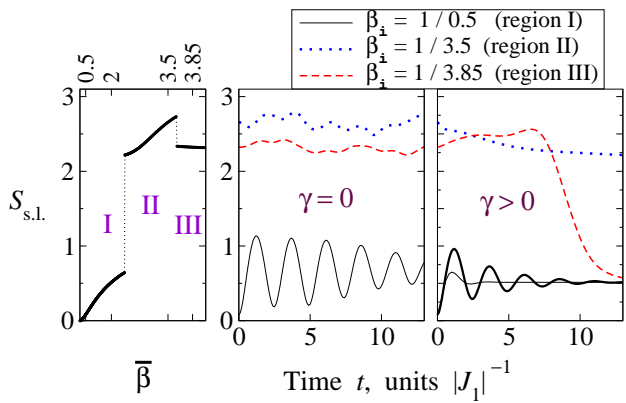}
\caption{\label{fig_time-evolution} 
Left panel shows sublattice entanglement entropy for $N=12$.  Center
[rightmost] panel shows evolution of $S_{\rm s.l.}$ after a quench, without
[with] dissipation.
Initial values are $\bar{\beta}_{\rm i}=0.5$ (solid), $\bar{\beta}_{\rm
i}=3.5$ (dotted), $\bar{\beta}_{\rm i}=3.85$ (dashed).
Final value is $\bar{\beta}_{\rm f}=2.0$ in each case.
Damping constant in rightmost panel is shown for $\gamma=0.5$ for all three
$\bar{\beta}_{\rm i}$.  For $\bar{\beta}_{\rm i}=0.5$, the thick solid line
has smaller damping, $\gamma=0.1$.
In the presence of dissipation, $S_{\rm s.l.}$ relaxes to the ground state
value when the initial state has same lattice momentum ($k=\pi$) but not when
the initial state has different lattice momentum ($k=0$).
}
\end{figure}

The fact that our system has true level crossings, place them in the complete
opposite limit to what is necessary for adiabatic quantum computing
\cite{AdiabaticQC}.  It is therefore perhaps natural to study the opposite
limit of adiabatic parameter changes, namely parameter \emph{quenches} where
$\bar{\beta}$ changes as a step function.
In an isolated finite spin ring, a quench of this type leads to oscillatory
behaviors of most quantities, including entanglement entropies for various
partitions; there is no relaxation mechanism for the system to reach its
ground state.  This is shown in Fig.~\ref{fig_time-evolution} (center panel).
Studying the evolution after a parameter quench from $\bar{\beta_{\rm i}}$ to
$\bar{\beta_{\rm f}}$ involves following the wavefunction
\[
\ket{\psi(t)} ~=~  \exp{\lbc-iH(\bar{\beta_{\rm f}})t\rbc} \; \ket{\psi(0)}
\]
explicitly in time, where the initial state $\ket{\psi(0)}$ is the ground
state of $H(\bar{\beta_{\rm i}})$.  We performed this calculation by expanding
the operator $e^{-iH(\bar{\beta_{\rm f}})t}$ to sufficiently high order for
each time step.

We also explore relaxation issues, specifically the effect of the
$0$-$\pi$-$0$-$\pi$- crossings, by adding an artificial dissipation to the
temporal evolution (Fig.~\ref{fig_time-evolution} rightmost panel).  Instead
of evolving in time $t$, we evolve in $t(1-i\gamma)$.  This form of damping
(`cooling') preserves the lattice momentum but not the energy.  We display
entanglement evolutions for the large-damping case of $\gamma=0.5$, and also a
single example with smaller damping, $\gamma=0.1$.
With damping, the sublattice entanglement $S_{\rm s.l.}(t)$ relaxes to the
ground-state $S_{\rm s.l.}$ corresponding to the final $\bar{\beta}$ only if
the starting state has the same lattice momentum $k$.
For $N=12$, two level crossings divides the parameter space into regions I and
III with the same $k=\pi$ and an intermediate region II with $k=0$.  The
entanglement entropy after a quench to $\bar{\beta}=2$ (region I) relaxes to
the `correct' value if the starting ground state is in region I or III, but
not if it is in region II.

The structure of level crossings and alternating ground-state momenta thus
allow selective targeting of quantum ground states, and therefore could serve
as the basis of selective variations of the basic idea behind adiabatic
quantum computing.

\section{Conclusions}

We have identified a mesoscopic system with unusual entanglement properties.
Our system is the finite-size analog of the intriguing
ferromagnetic-antiferromagnetic chain.
The surprising long-distance concurrence we have found, with tunable distance,
makes this system ideal for various quantum information processing tasks.  For
example, \emph{entanglement transport} is here simply a matter of tuning the
interaction parameter while remaining in the ground state.
Our calculations of multi-qubit entanglement entropies and the fidelities have
revealed unexpected aspects of the level crossings present in these systems.
We have also presented entanglement dynamics results, again indicating
possible quantum information processing applications. 

Our work opens up a number of questions.  First, our focus on the
$0$-$\pi$-$0$-$\pi$- level crossings raises the question: which of the
entanglement behaviors reported here also carry over to the other systems
where such level crossings have been observed?  To the best of our knowledge,
no unified study of such crossings exists.  Our entanglement results call for
a general investigation of finite-size level crossing sequences occurring as a
result of incommensurate correlations.

Second, our study of entanglement dynamics treats only a minute fraction of
the possible kinds of dynamics, and only a single entanglement measure out of
many.  Since entanglement measures are sensitive to a variety of wavefunction
characteristics not necessarily present in traditional condensed-matter
measures such as correlation functions, the dynamics of these new measures
present rich opportunities for characterizing non-equilibrium phenomena.
Again, comprehensive investigations are called for, but the study of the
evolution of various entanglements would be appropriately first done in
simpler (perhaps exactly solvable) many-body systems.

\acknowledgments

We thank A.~L\"auchli, J.~Richter, and N.~Shannon  for useful discussions.

\end{document}